# Gate-Tunable Spin-Orbit Coupling in a Germanium Hole Double Quantum Dot

He Liu,[1,2,#] Ting Zhang,[1,2,#] Ke Wang,[1,2,#] Fei Gao,[3,] Gang Xu,[1,2] Xin Zhang[1,2], Shu-Xiao Li,[1,2] Gang Cao,[1,2] Ting Wang,[3] Jianjun Zhang,[3] Xuedong Hu[4], Hai-Ou Li,[1,2,*] and Guo-Ping Guo[1,2,5,*]

[1] *CAS Key Laboratory of Quantum Information, University of Science and Technology of China, Hefei, Anhui 230026, China*

[2] *CAS Center for Excellence and Synergetic Innovation Center in Quantum Information and Quantum Physics, University of Science and Technology of China, Hefei, Anhui 230026, China*

[3] *Institute of Physics and CAS Center for Excellence in Topological Quantum Computation, Chinese Academy of Sciences, Beijing 100190, China*

[4] *Department of Physics, University at Buffalo, SUNY, Buffalo, New York 14260, USA*

[5] *Origin Quantum Computing Company Limited, Hefei, Anhui 230026, China*

[#]These authors contributed equally to this work

* Corresponding author. Emails: haiouli@ustc.edu.cn (H.-O. L.); gpguo@ustc.edu.cn (G.-P.G.).

Hole spins confined in semiconductor quantum dot systems have gained considerable interest for their strong spin-orbit interactions (SOIs) and relatively weak hyperfine interactions. Here we experimentally demonstrate a tunable SOI in a double quantum dot in a Germanium (Ge) hut wire (HW), which could help enable fast all-electric spin manipulations while suppressing unwanted decoherence. Specifically, we measure the transport spectra in the Pauli spin blockade regime in the double quantum dot device. By adjusting the interdot tunnel coupling, we obtain an electric-field-tuned spin-orbit length $l_{SO}$ = 2.0–48.9 nm. This tunability of the SOI could pave the way toward the realization of high-fidelity qubits in Ge HW systems.

# I. INTRODUCTION

Hole spin qubits in Ge quantum dots (QDs) are intriguingly attractive in quantum information processing because of their advantageous properties [1-8]. Compared with the III-V materials, natural Ge—a group-IV semiconductor—contains a much lower abundance of nuclear-spin isotopes, and can be further purified to become a nuclear-spin-free host, greatly improving the coherence times of spin qubits [9,10]. Furthermore, because the hole states are formed of p-atomic-orbital wave functions, the contact hyperfine interaction (HFI) vanishes completely, and the anisotropic HFI (dipole-dipole and angular momentum terms) dominates [11]; the latter can be reduced or even eliminated by motional averaging [12-15]. Another significant advantage of hole spins over electron spins is their much stronger spin-orbit interaction (SOI) [16], which allows an all-electrical manipulation of single hole spins [1-8,17-20] and simplifies device fabrication. As a material platform, Ge hut wires (HWs) [21] are considered a strong contender for large-scale quantum circuits because of some favorable properties. For example, the direct Rashba SOI (DRSOI) [22,23] in one-dimensional hole nanowires has yielded the fastest Rabi frequency to date [3], and the realization of site-controlled Ge HWs [24] could benefit the scaling up of spin qubit applications.

The main challenge faced by a hole-spin-based quantum computer lies in the extremely strong DRSOI in Ge HWs. On the one hand, it is the crucial ingredient that allows ultrafast qubit operations. On the other hand, it enables undesirable decoherence by allowing strong coupling to all electrical fluctuations in the environment, such as phonons and charge noise [25-32]. To benefit from DRSOI while overcoming its drawbacks, one can take advantage of the fact that DRSOI is highly tunable by electric fields [22,23]. One can thus realize different features (e.g., fast operation speeds and long coherence times) in different SOI strengths, adjusted by gate voltages [33]. Such tunability has indeed been explored recently in experiments demonstrating a spin-orbit switch in a core-shell Si/Ge nanowire [2].

One interesting feature of the Ge HW system is the variety of SOI it has. In addition to the DRSOI, other SOI mechanisms such as interface SOI and intrinsic SOI may also be present and important [34,35]. In particular, by applying a certain electric field, we may be able to completely turn off the total SOI at an operating sweet spot for the qubits (idle state), where the effects of charge noise and phonons are strongly suppressed [34]. This electrical tunability of the SOI could also provide a means for precise control over the g-factor, which can be exploited to address qubits individually in a large-scale array [33]. Therefore, a better understanding and control of the SOI in Ge HW QDs is of critical importance if longer relaxation and coherence times and higher-quality spin control are to be gained.

In this work, we determine and study the main hole spin relaxation mechanisms on a highly tunable double quantum dot (DQD) fabricated in a Ge HW. We measure the leakage current in the Pauli spin blockade (PSB) regime to probe the various relaxation processes in the DQD by varying the applied magnetic field and the interdot detuning. By increasing the tunnel coupling within two dots, we find that the current peak is split into two peaks in a magnetic field. This transition is induced through the combined effects of SOI and Zeeman splitting. Based on our numerical simulations [36], we extract a gate-tunable SOI strength with spin-orbit length $l_{SO}$ = 2.0–48.9 nm, which is a clear evidence for a potential spin-orbit switch in a Ge HW DQD.

## II. RESULTS AND DISCUSSIONS

### A. Experimental setup

Our few hole DQD device (see Appendix for more details) was fabricated following established procedures by gating a Ge HW [Fig. 1(a)]. The self-assembled Ge HWs are grown on a Si(001) wafer by solid-source molecular beam epitaxy [21]. After wet etching with buffered hydrofluoric acid, two 30-nm-thick palladium contact pads with a gap of around 265 nm are formed by electron beam evaporation. The sample is then covered with a 20 nm insulating layer of hafnium oxide to suppress gate leakage. Finally, three 35-nm-wide top gates are fabricated with titanium/palladium (3/25 nm thick). We performed the measurements in a liquid He-3 refrigerator at a base temperature of 240 mK, with an out-of-plane magnetic field.

By applying positive voltages to three top gates to create the confinement potential, a DQD in series along the nanowire is formed between them [Fig. 1(c)]. In the Coulomb blockade region, the number of hole occupations (m, n) in the left and right dots can be adjusted using gates L and R; gate C is used to adjust the tunnel coupling between the two dots. Fig. 1(b) shows a typical charge stability diagram of the DQD measured from transport current with a source-drain bias $V_{SD} = +2.5$ mV. With hole transport only allowed at triple points, we detect an array of bias triangles. Here the green lines separating the different Coulomb blockade regions indicate the change of hole occupations in the DQD, denoted by (m, n). From the slopes and spacings of the transition lines, we obtained the values of the lever arms and charging energies for each QD (see APPENDIX).

## B. Pauli spin blockade

The bias triangle in the red dashed circle in Fig. 1(b) exhibits PSB characteristics of particular interest to us [37-39]. PSB normally occurs in the transition from charge state (2m+1,2n+1) to (2m,2n+2), which can be equivalently described as (1,1) to (0,2) for simplicity and thought of as a feature of the two-hole spectrum. When the DQD is in the T(1,1) state [Fig. 2(a)], transport is blocked: the transition from T(1,1) to S(0,2) is forbidden by spin conservation, while the energy of T(0,2) is too high to access [37].

Figure 2(b) shows the zoom-in of the bias triangle marked in Fig. 1(b) by the red dashed circle. Because of PSB, we observe current rectification in the trapezoidal region indicated by the dashed line. We modified the detuning of the potentials of the left and right QDs by changing the voltages of gates L and R (red arrow). Once the energy levels of T(1,1) and T(0,2) align, the PSB is lifted and transport through the DQD is allowed, leading to an enhanced current at the top of the bias triangle. When we reverse the source-drain bias, a non-zero current is observed throughout the triangular region in Fig. 2(c) because the hole can transition freely from the S(0,2) to the S(1,1) state. In addition, after applying a 50 mT magnetic field, a large leakage current appears at the base of the triangle in Fig. 2(d), which is induced by the SOI and on which we will elaborate next.

### C. Transport spectra induced by the SOI

The PSB is lifted when a spin is flipped. Such spin transition could occur via a variety of mechanisms: i) the SOI can hybridize T(1,1) with S(0,2) via spin-flip tunneling [40,41]; ii) the HFI can mix the different (1,1) states [42,43]; iii) spin-flip cotunneling to the leads allows charge and spin exchange with the leads [44-47]; and iv) the difference in g-factors between the two dots couples $T_0(1,1)$ and S(1,1) states [36], similar to the longitudinal Overhauser field, giving rise to a finite leakage current in the PSB region. To identify and study the main spin-relaxation mechanism(s) in our system, we measured the leakage current spectra as a function of detuning and magnetic field with different interdot tunnel couplings in the bias triangle of Fig 2(b). Varying the gate C voltages could adjust the tunnel coupling between the two dots over a wide range, giving us a tuning knob to differentiate mechanism i) from the others.

In the PSB regime, we find that the measured leakage current spectra with relatively strong and weak interdot tunnel coupling [Figs. 3(a) and 3(b), respectively] show two completely different field-dependent behaviors. In the strong tunnel coupling regime ($V_C = 596$ mV), the current spectrum shows a double-peak structure consisting of a dip at zero field and two current peaks at finite magnetic fields. This deep zero-field dip and the experimental temperature of 240 mK exclude the mechanism of spin-flip cotunneling which leads to only a shallow dip for $k_B T \ll t$ [47]. Furthermore, spin mixing induced by HFI is effective at the magnetic field of a few milli-Tesla because of the weak HFI in Ge hole systems [11,48]. Therefore, the field dependence here can be well explained by a strong SOI and Zeeman splitting [see energy level diagrams in Figs. 3(c) and 3(d)]. At $B = 0$ mT, the tunnel coupling between the two dots couples S(1,1) to S(0,2) with coupling strength *t*, leading to strong hybridization and a large level anticrossing [Fig. 3(c)], while the three degenerate triplets T(1,1) do not couple with S(0,2) through spin-preserving tunneling, and are energetically detuned from the singlet states so that the effect of spin-flip tunneling is minimal. Furthermore, the large singlet anticrossing also limits the hyperfine mixing between the (1,1) states, suppressing the leakage current at zero field [42]. However, at a finite magnetic field, the strong SOI in Ge HWs hybridizes $T_\pm(1,1)$ and $S(0,2)$ with coupling strength $t_{SO}$ [Fig. 3(d)], thereby increasing the leakage current. Given the distance between current peaks induced by the SOI scales with strength $t$, in the weak tunnel coupling regime ($V_C = 612$ mV), we only observe a single peak centered at zero field with a width of approximately 300 mT, which excludes any HFI effect [42,48]. In the strong tunnel coupling regime, on the other hand, the large singlet anticrossing helps separate the two SOI-induced peaks in the leakage current, so that we can observe both of them experimentally. Using the slopes of the observed current lines [as denoted by the yellow dashed line in Fig. 3(b)], which can be interpreted as indicating resonances between $T_-(1,1)$ and $S(0,2)$, we obtain g = 3.17, which is in agreement with values obtained previously [49].

To better understand and quantitatively describe the observed transition, a series of line cuts [Fig. 3(e)] around zero detuning ($\varepsilon = 0$) were analyzed in the leakage current spectra [as denoted by the purple dashed line in Fig. 3(a)] for different tunnel couplings. With decreasing tunnel coupling between the two dots, specifically, varying gate C voltages from 596 mV to 612 mV, the magnetic field dependence of the leakage current induced by the SOI eventually transforms from a double-peak structure to a single peak. In this process, the magnitude of the leakage current and the distance between the two current peaks gradually decrease. Consequently, the dip at zero field finally disappears or is too narrow to be probed. In a high magnetic field, the energy level of $T_-(1,1)$ is pushed below $S(0,2)$ and drives the system into a Coulomb blockade regime that suppresses the current. This SOI-induced behavior in the leakage current that we observed is similar to that in Ge/Si core/shell nanowires where it is measured in two different bias triangles [36]. Following Ref. [36], we applied the same modified model which considers the effects of both SOI and the difference in g factors in the two dots to analyze the leakage current data.

## D. Theoretical model and simulation

The bias triangle we study here is in the few-hole regime [3]. The Hamiltonian matrix of our effective two-hole system in the Pauli blockade regime can then be presented on the triplet-singlet basis [$T_+(1,1)$, $T_-(1,1)$, $T_0(1,1)$, $S(1,1)$, $S(0,2)$] as [36]

$$H = \begin{pmatrix} E_Z & 0 & 0 & 0 & it_+ \\ 0 & -E_Z & 0 & 0 & it_- \\ 0 & 0 & 0 & \xi B & 0 \\ 0 & 0 & \xi B & 0 & t \\ -it_+ & -it_- & 0 & t & -\varepsilon \end{pmatrix}. \tag{1}$$

Here, $E_Z$ describes the energy shift of polarized triplets with respect to the unpolarized triplet in a magnetic field. $\varepsilon$ is the detuning between S(1,1) and S(0,2) states and is set to zero. $t$ is the spin-preserving tunneling matrix element between S(1,1) and S(0,2) states. $t_+$ and $t_-$ are the spin-flipping tunneling matrix elements induced by the SOI from the two polarized triplets to S(0,2), which satisfy the relation $t_+ = -t_- = t_{SO}$ due to time-reversal symmetry [41]. $\xi B$ describe the mixing between $T_0(1,1)$ and S(1,1), which is caused by the difference of g-factors in the two dots, with $\xi = \frac{1}{2}(g_L - g_R)/(g_L + g_R)$ [36,50]. The value of the effective g-factor is expected to be site-dependent and depends on the microscopic characteristics of QDs, such as the confining potential [22,51], charge occupation [49], and the wave function [52]. We estimate that the value of $\xi$ is around 0.14 from the electric-dipole spin resonance spectra of similar samples in Ge hut wires [3].

We diagonalize the above Hamiltonian and describe the dynamics of hole transport in our system with a master equation in the new basis

$$\frac{d\rho}{dt} = -i[H^{\text{diag}}, \rho] + \Gamma\rho + \Gamma_{\text{rel}}\rho. \tag{2}$$

Here, $\rho$ is the density matrix. $\Gamma$ describes the hole transport between the DQD and leads including decay to the drain lead and reload from the source lead. $\Gamma_{\text{rel}}$ describes the relaxation process from excited states to the ground state in the DQD. We obtain the density matrix of the steady-state by solving the equation $\frac{d\rho}{dt} = 0$, and the current through the DQD can be expressed as $I = \sum_n p_n \Gamma |\langle n|S(0,2)\rangle|^2$, where $p_n = \rho_{nn}$ and $|n\rangle$ refer to the eigenstates of Hamiltonian (1).

The theoretical model is particularly successful at reproducing the zero-detuning current traces [Fig. 3(e), solid lines] with parameters listed in Table I. Through the numerical simulation, we find that the magnitude of the leakage current is directly related to the tunneling rate between the DQD and the lead, $\Gamma$, and the relaxation rate $\Gamma_{\text{rel}}$ directly determines the depth of the dip (i.e. the value of current at zero field). The tunneling rates within two dots ($t$ and $t_{\text{SO}}$) cannot be determined independently and are related to the specific shape of the current curve.

### E.  Tunability of the SOI

Given the results of model parameters $t$ and $t_{\text{SO}}$ (see Table I), we extracted a spin-orbit length of $l_{\text{SO}} = 2.0$–$48.9$ nm from $\frac{t_{\text{SO}}}{t} = \frac{4}{3}\frac{l_{\text{dot}}}{l_{\text{SO}}}$ [53,54] and a dot-to-dot distance $l_{\text{dot}}$ of ~75 nm. $\xi$, the relative g-factor difference between the two dots, is an assumed parameter in the calculation of $l_{\text{SO}}$. During the fitting process, we found that when the value of $\xi$ varies within a certain range, the fitted value of $t_{\text{SO}}/t$ hardly changes, which gives us confidence that the calculated value of $l_{\text{SO}}$ should be reliable.

Here, the distance between two dots remains almost unchanged when adjusting the voltage of gate C with the same hole occupation, which we have verified via simulation with COMSOL. This is a direct evidence that the spin-orbit length in our system could be highly tunable by changing gate C voltage [Fig. 4(a)]. A smaller $l_{\text{SO}}$ at higher gate voltages is reasonable and is consistent with the known relation between the direct Rashba coefficient and the average electric field tunable by the gate voltage [22,23]. These remarkably short $l_{\text{SO}}$ we obtained indicate a strong DRSOI in our system, and the tunability of the spin-orbit length is consistent with the results obtained in Ge/Si core/shell nanowires [2]. In order to extend the tunable range of $l_{\text{SO}}$ further, we would need to optimize the electrode design to decrease the strength of the electric field, or directly reduce the gate C voltage according to the curve trend in Fig. 4(a), resulting in weaker SOI.

The high degree of tunability of $l_{SO}$ is the key to a spin-orbit switch, which can be used to enable fast Rabi oscillation with strong SOI, and ensure longer coherence times when decreasing SOI [3]. This switchable qubit operation scheme breaks the trade-off between coherence and speed of control (i.e. Rabi frequency), and greatly improves the qubit fidelity theoretically, which is of critical importance for fault-tolerant quantum computing [55]. Considering the presence of Dresselhaus SOI induced by the interface inversion asymmetry in Ge HWs [35], a 'sweet spot' of operation may be present when the total SOI is completely turned off, as shown in Fig. 4(b) [34]. Here, the effective spin-orbit fields caused by different SOI mechanisms ($B_R$ and $B_D$) are represented by arrows of different colors. The red arrow indicates the total spin-orbit field and is quenched when $B_R$ and $B_D$ are equal in amplitude but opposite in direction (marked in red star), which can be realized by combining the two characteristic advantages in Ge HWs: i) the tunability of the DRSOI; ii) the site-controlled growth mode of nanowires [24].

Our results are obtained in the few-hole regime, though we expect that SOI strength should remain highly tunable in quantum dots in the single-hole regime, as theoretical studies of DRSOI [22,23] and other forms of SOI [33-35] were all done for single hole quantum dots and did not require the presence of extra electrons or holes. In the meantime, applications in quantum information processing may even be easier in the multi-hole regime [1-3,19], as have been explored before for multi-electron spin qubits both theoretically and experimentally [56-59]. As such, our study of tunable SOI should be relevant in a variety of situations.

## III. CONCLUSION

In summary, we demonstrate experimentally a strongly tunable spin-orbit interaction in a lithographically defined DQD in a Ge hut wire that exhibits excellent charge stability. The multi-gate architecture provides independent control of the electron number in each dot as well as a highly tunable tunnel coupling from 0.4 to 11.5 μeV. By studying the magnetic field dependence of leakage current in the PSB regime, we identified SOI as the dominant spin-relaxation mechanism in our system. With increasing tunnel coupling, we observed a transition in the leakage current from a single peak at zero field to two peaks at finite magnetic fields induced through the effect of SOI. Numerical calculations yield quantitative agreement with experimental results, showing a strong and tunable SOI with spin-orbit length $l_{SO}$ = 2.0–48.9 nm in our system. These results are promising evidences for potential spin-orbit switches and high-fidelity qubits in Ge HW QDs.

## Acknowledgments

This work was supported by the National Key Research and Development Program of China (Grant No.2016YFA0301700), the National Natural Science Foundation of China (Grants No. 12074368, 92165207, 12034018, 11804315, and 61922074), the Anhui Province Natural Science Foundation (Grants No. 2108085J03), the USTC Tang Scholarship. X. H. acknowledges financial support by U.S. ARO through No. W911NF1710257, and this work was partially carried out at the USTC Center for Micro and Nanoscale Research and Fabrication.

## APPENDIX: MEASUREMENT DETAILS

Figure 5(a) shows the network of capacitors and voltage nodes that are used to calculate the conversion factors between the gate voltages and energy. Here we consider the cross-talk between gate L (R) and dot R (L) and the four $\alpha_{il}$s describe the coupling of the gate $i$ to the energy offset of their respective dot $l$. Fig. 5(b) shows a charge stability diagram over a wider range of gate sweeping compared to the one in Fig. 1(b). Notice the irregular distance between current peaks here, which is an indication that the addition energy includes both Coulomb interaction and single-particle excitation, and the system is in the few-hole regime. The gradient [$k$ in Fig. 5(b)] of the charge transition (fit in yellow) yields the relative effect of the two gates on the single-particle energy offset of the same dot.

$$k_R = \frac{\delta V_L}{\delta V_R} = -\frac{\alpha_{RR}}{\alpha_{LR}} = -13 \tag{1}$$

$$k_L = \frac{\delta V_L}{\delta V_R} = -\frac{\alpha_{RL}}{\alpha_{LL}} = -\frac{1}{22} \tag{2}$$

Figure 5(c) shows the bias triangle with a source-drain voltage $V_{SD} = +2.5$ mV at the magnetic field of 50 mT. As the difference between the single-particle energies of two dots stays fixed along a polarization line (the base of the triangle), we can determine the relative weights of the $\alpha_{il}$s from the gradient ($k_p$) of this line in Fig. 5(c).

$$\alpha_{RR}\delta V_R + \alpha_{LR}\delta V_L = \alpha_{LL}\delta V_L + \alpha_{RL}\delta V_R \tag{3}$$

$$k_p = \frac{\delta V_L}{\delta V_R} = 0.77 \tag{4}$$

At last, the absolute value of $\alpha_{il}$ which is called the lever arm as well can be found from the length of the base of the triangle.

$$\alpha_{RR}\Delta V_R + \alpha_{LR}k_p\Delta V_R = eV_{SD} = 2.5 \text{ meV} \tag{5}$$

Here $\Delta V_R$ denotes the voltage change of gate R in the range of the bias triangle. Using the equations from (1) to (5), we can extract the four values of the lever arms between gate voltages and energy of dot $\alpha_{LL} = 384.1$ meVV$^{-1}$, $\alpha_{LR} = 22.7$ meVV$^{-1}$, $\alpha_{RR} = 295$ meVV$^{-1}$, $\alpha_{RL} = 17.5$ meVV$^{-1}$. Then we can obtain the charging energies of two dots from the spacing of the charge addition lines as shown in Fig. 5(d). $E_C^R = \alpha_{RR}\Delta V_R = 8.6$ meV, $E_C^L = \alpha_{LL}\Delta V_L = 5.2$ meV.

**Figure Captions:**

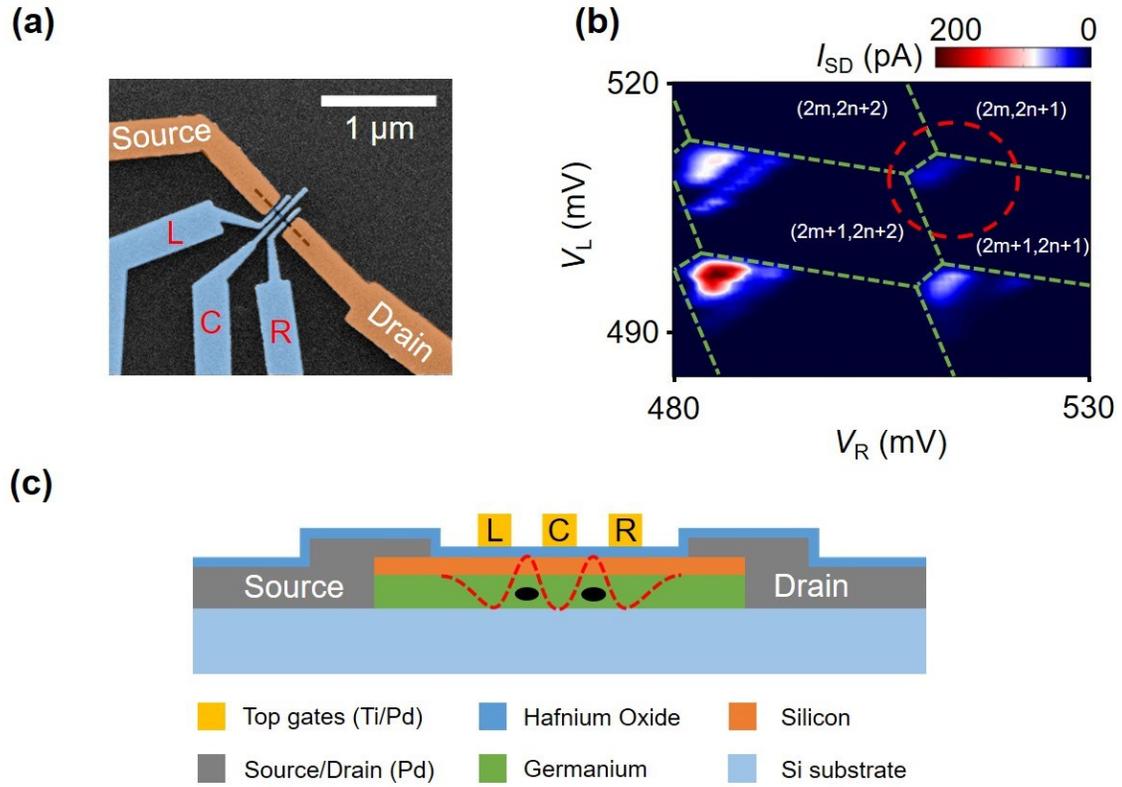

FIG. 1. (a) Scanning electron microscopy image of the hole DQD in a Ge HW. (b) Charge stability diagram of the Ge DQD as a function of $V_L$ and $V_R$ with $V_{SD} = +2.5$ mV. Green dashed lines separate the charge states. The bias triangle in the encircled region is investigated in detail. (c) Schematic cross-section structure of the Ge HW DQD.

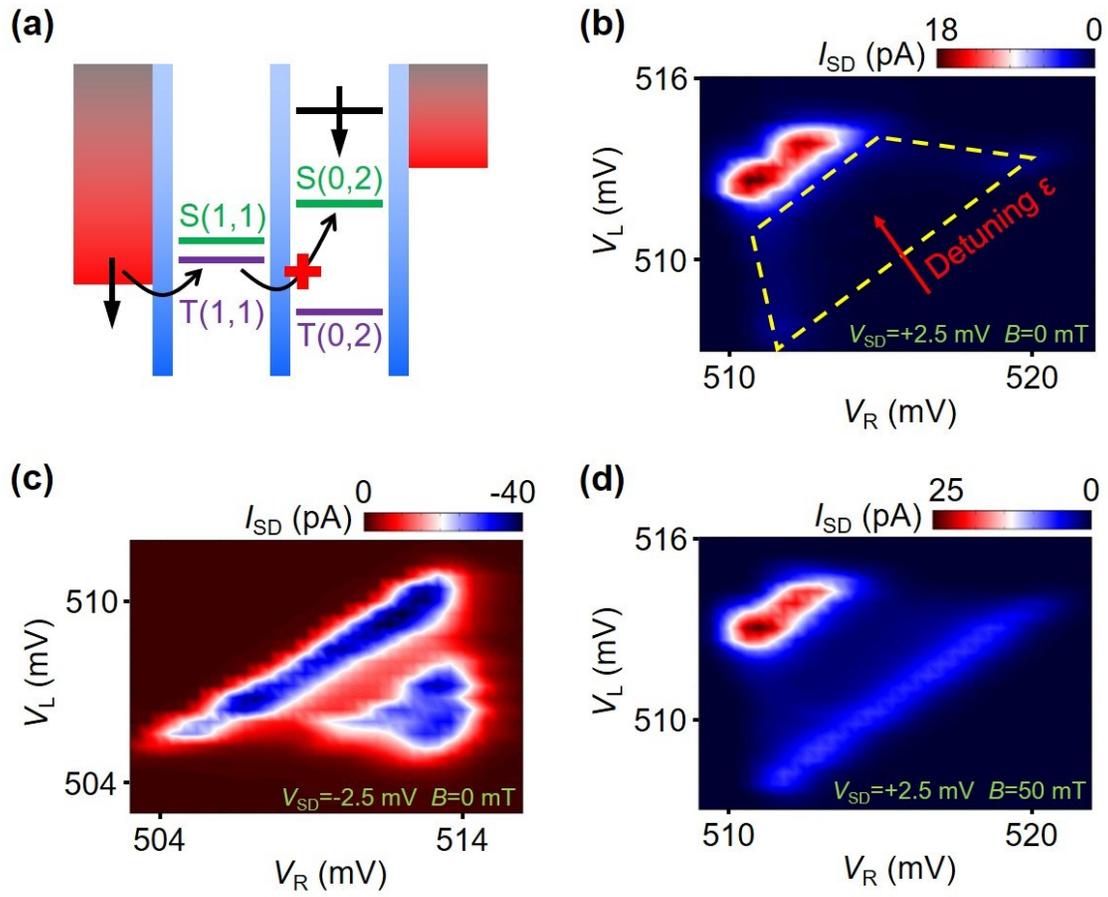

FIG. 2. (a) Schematic diagram of the PSB for a hole DQD. (b) The bias triangle described in the main text exhibits the PSB signatures. Strong suppression of the current is observed in the region bounded by the dashed line. Detuning of DQD was changed along the red arrow. (c) Reversing the bias, an enhancement of the leakage current is observed throughout the triangle. (d) At positive bias, a large leakage current appears at the base of the triangle with a 50 mT magnetic field due to SOI. Here, the voltage of gate C is 596 mV.

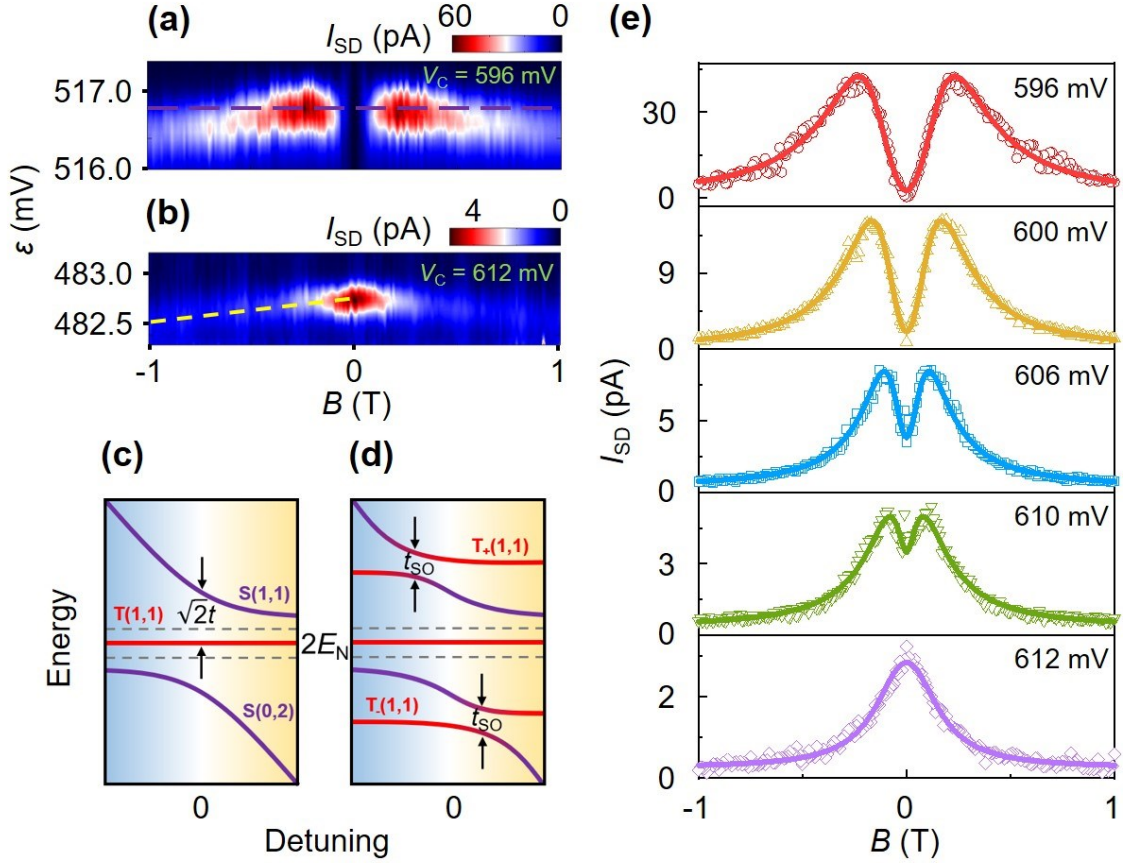

FIG. 3. (a,b) Leakage current spectra induced by the SOI with different tunnel coupling for magnetic fields in the range (−1 T–1 T). For large tunnel couplings in (a) ($V_C = 596$ mV), the spectrum shows a double-peak structure that consists of a dip at zero field and two current peaks at specific magnetic field strengths. For small tunnel coupling in (b) ($V_C = 612$ mV), the spectrum shows a single peak at zero magnetic field. (c,d) Schematic energy-level diagrams for large tunnel couplings needed to explain the behavior of leakage current in Fig. 2(a). (e) Series of line cuts near zero detuning [as denoted by the purple dashed line in (a)] with different tunnel couplings. Increasing gate C voltage from 596 mV to 612 mV decreases the coupling strength. The solid lines are fitted curves using the modified model of Ref. [36].

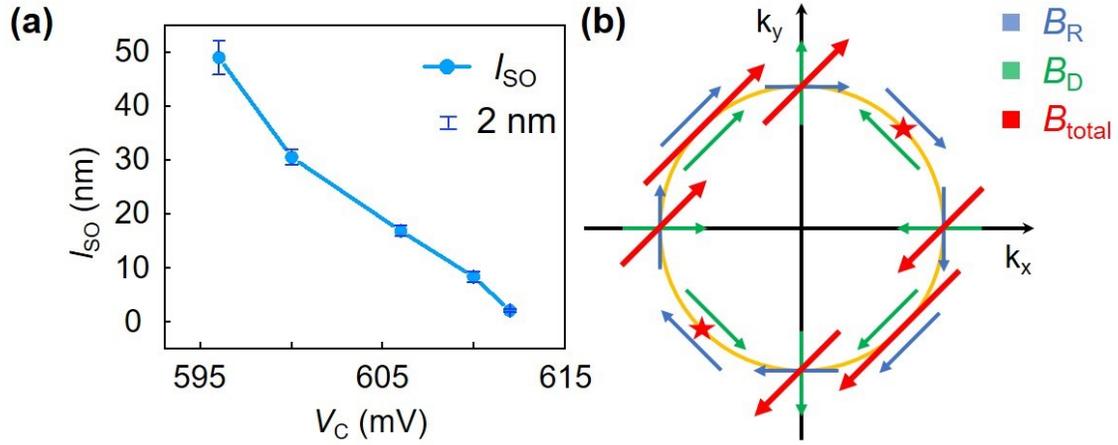

FIG. 4. (a) Spin-orbit length as a function of gate C voltage, as extracted from fits to line cuts in Fig. 3(e). The error bars represent the calculation error of $l_{SO}$. They are determined by the fitting errors of parameters $t$ and $t_{SO}$ at difference $V_C$. (b) In momentum space, the effective spin-orbit fields are denoted by arrows of different colors: blue, direct Rashba SOI field ($B_R$); green, Dresshaus SOI field ($B_D$); red, total SOI field ($B_{total}$). $B_R$ and $B_D$ are equal in amplitude, and the total spin-orbit field will be completely turned off at the points marked by the red star.

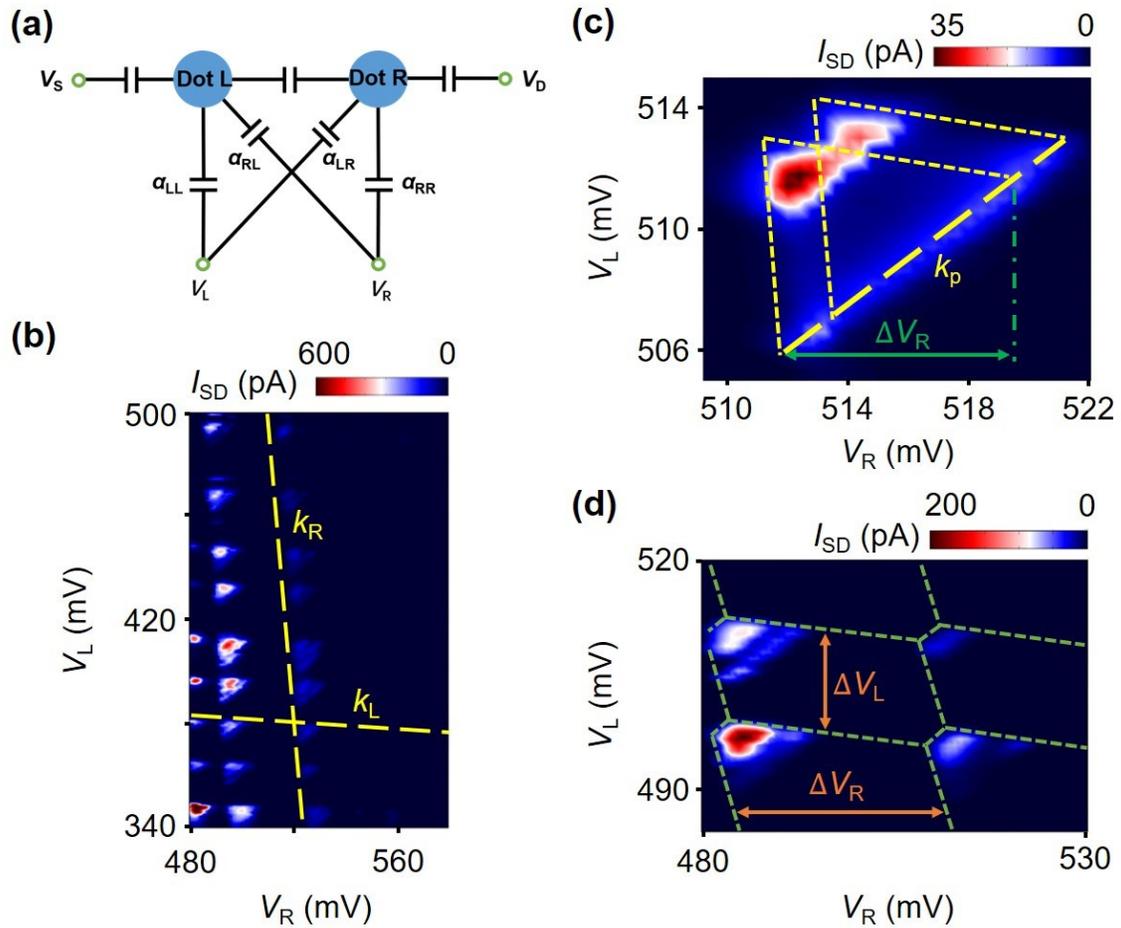

FIG. 5. (a) The network of capacitors and voltage nodes that are used to calculate the level arm. (b) The charge stability diagram of the DQD over a wide range of gate voltages. The yellow dashed lines denote the charge addition lines of two dots. (c) The bias triangle with a source-drain voltage $V_{SD} = +2.5$ mV at the magnetic field of 50 mT. $k_p$ is the gradient of the baseline of the triangle. (d) The same charge stability diagram as shown in Fig. 1(b). $\Delta V_i$ is the spacing of the charge addition line of dot $i$.

**Table Captions:**

| Data | | | Assumed parameter | Fitted parameters | | | | | Calculation |
|---|---|---|---|---|---|---|---|---|---|
| $V_C$ (mV) | $V_L$ (mV) | $V_R$ (mV) | $\xi$ | $\Gamma$ (MHz) | $\gamma$ | $t$ (μeV) | $t_{SO}$ (μeV) | $I_0$ (pA) | $l_{SO}$ (nm) |
| 596 | 510 | 516 | 0.14 | 1320±20 | 0.008±0.001 | 11.5±0.5 | 23.5±0.5 | 0.25±0.05 | 48.9±3.2 |
| 600 | 501 | 500 | 0.14 | 510±5 | 0.02±0.003 | 6±0.2 | 19.7±0.3 | 0.05±0.05 | 30.5±1.5 |
| 606 | 501 | 494 | 0.14 | 335±5 | 0.055±0.003 | 2.6±0.1 | 15.5±0.3 | 0.4±0.05 | 16.8±1.0 |
| 610 | 502 | 490 | 0.14 | 235±5 | 0.075±0.003 | 1.2±0.1 | 14.5±0.5 | 0.35±0.05 | 8.3±1.0 |
| 612 | 499 | 483 | 0.14 | 38±1 | 2.1±0.3 | 0.4±0.05 | 20±1 | 0.25±0.03 | 2.0±0.4 |

TABLE I. Fit parameters for the leakage current induced by SOI. $\Gamma$ describes the tunneling rate of S(0,2) to drain lead. $\gamma = \Gamma_{\text{rel}}/\Gamma$, here $\Gamma_{\text{rel}}$ describes the rate of relaxation from excited states to the ground state. $t$ describes the tunnel coupling between the dots and $t_{SO}$ describes the coupling between the triplet T(1,1) and singlet S(0,2). $I_0$ is the background current. We can extract the spin-orbit length $l_{SO}$ from the relation $\frac{t_{SO}}{t} = \frac{4}{3}\frac{l_{\text{dot}}}{l_{SO}}$.